\documentclass[twocolumn,showpacs,preprintnumbers]{revtex4}
\usepackage{dcolumn}
\usepackage{bm}
\input epsf

\begin{document}

\preprint{GACG-07-04}

\preprint{astro-ph/0701624}

\title{Inflation, Reheating and Dark Matter}

\author{V\'{\i}ctor H. C\'{a}rdenas}
\email{vcardenas@unab.cl}

\affiliation{Departamento de F\'{\i}sica y Matem\'atica,
Universidad Andres Bello, Los Fresnos 52, Vina del Mar, Chile}

\begin{abstract}
In a recent paper, Liddle and Urena-Lopez suggested that to have a
unified model of inflation and dark matter is imperative to have a
proper reheating process where part of the inflaton field remains.
In this paper I propose a model where this is possible. I found that,
incorporating the effect of plasma masses generated by the inflaton
products, enable us to stop the process. A numerical estimated
model is presented.

\end{abstract}


\maketitle

\section{Introduction}

In a recent paper \cite{LUL}, Liddle and Ure\~na-Lopez studied the
conditions under which we can have a unified description of
inflation \cite{reheat}, dark energy \cite{DE} and dark matter \cite{DM}.
The key ingredient required is to have a reheating process where not
all the inflaton energy density decays into radiation. The crucial role of
reheating in these type of models was found earlier in \cite{VHC},
where models of quintessential inflation were proposed. The
authors of \cite{LUL} found that the standard reheating mechanism
\cite{reheat} can not be used for this task, because in this
scenario the scalar field decays completely. They also studied the
preheating scenario \cite{preheat}, where it is known that the
field decay can be incomplete, however as was discussed in
\cite{LUL} the simplest model based on a theory with a quadratic
scalar field turns out inconsistent with observations.

In this work I revisited this issue. Specifically I am interested
in a unified description of inflation and dark matter. The key
problem is to consider an appropriate phase of reheating that both
serve as a bridge between inflation and the standard model of
cosmology, and also give us an observationally consistent amount
of dark matter. I found that even using the standard
(perturbative) reheating mecanism, we can obtain a partially
decaying inflaton field. However, this observation does not help
to build the model by itself, because for a large initial
amplitude of the inflaton field, which is $\sim M_p$ at the end of
inflation, we can not neglect the parametric resonance effects
\cite{preheat}. So, we have to consider the preheating scenario
\cite{KLS97}, in which the first stages of reheating occur in a
regime of a broad parametric resonance, then the resonance becomes
narrow, and the last stage can be described by the elementary
theory of reheating \cite{reheat}. However, this time the last
stage is not so simple as the standard one, because the particles
created at the first stages affects the evolution of the inflaton
field. One way to consider this, is take into account the
generation of plasma masses for the inflaton decays products
\cite{KNR}, which can stop the particle creation process, doing it
kinematically forbidden. In section II we describe the basics of
the model we studied. The observational constraint is discussed in
section III, and applied to the model in section IV where is made
manifest the problem. The solution is described in section V where
the plasma masses are taking into account.

\section{The model}

Let us assume a theory with a quadratic scalar potential $V(\phi)=
V_0 + m^2\phi^2/2$. Here $V_0$ is a small positive constant needed
to explain dark energy \cite{DE}. During inflation, the friction
produced by the expansion of the universe makes the field evolve
slowly towards its vacuum, e.g. $m \ll H$. In this case the
equations controlling the evolution are
\begin{equation}
H^2 \simeq \frac{4\pi m^2}{3 M_p^2}\phi^2, \hspace{1cm}
3H\dot{\phi}+m^2\phi \simeq 0.
\end{equation}
Inflation last until the kinetic energy of the field equals the
potential energy $\dot{\phi}^2 \simeq V(\phi)$. As is well known
\cite{reheat} the field at the end of inflation takes the value
$\phi \simeq M_p$. Of course the end of inflation coincides with
the condition $m \simeq H$, as can be seen clearly from the
Friedman equation. After inflation the universe enters into the
reheating phase, the process where almost all the particles in the
universe were created. During this phase the scalar field
continues rolling down the hill of the potential towards its
minimum and starts to oscillate around it. In numerical estimates
one realizes that during the first oscillations the expansion of
the universe is still important \cite{KLS97} and the amplitude of
the field falls down very quickly (see Figure 1).
\begin{figure}[h]
\centering \leavevmode\epsfysize=8cm \epsfxsize=9cm\epsfbox{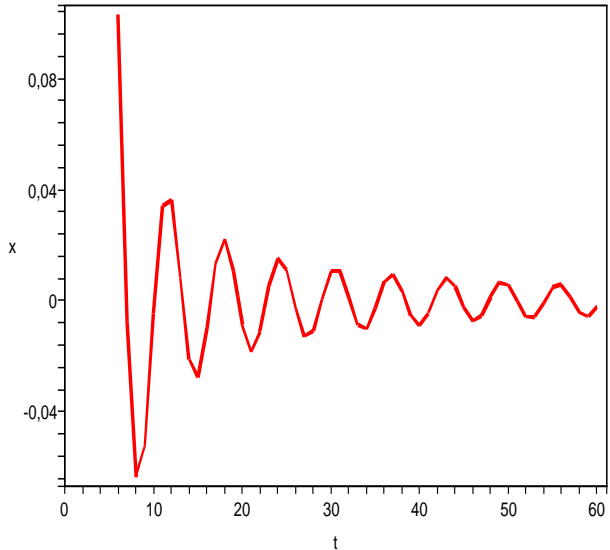}\\
\caption[fig1]{\label{fig1} Oscillations of the inflaton field
after inflation. The value of the scalar field is measured in
units of $M_p$, and time in units of $m^{-1}$.}
\end{figure}
After the first oscillation, the amplitude reaches the value $0.04
M_p$, indicating that the expansion is still important. Later the
scalar field enters into the oscillatory regime where the
amplitude decreases slowly
\begin{equation} \label{2}
\phi(t)= \frac{M_p}{\sqrt{3\pi}mt}\sin(mt)=\Phi(t)\sin(mt).
\end{equation}
It is during this stage where the average over many oscillations of
the scalar field can be described as non-relativistic matter
$\rho_{\phi} \simeq a^{-3}$. Reheating occurs when the amplitude
of the field decreases more rapidly than Eq.(\ref{2}).
Historically this process was studied first introducing an {\it
ad-hoc} term in the equation of motion for the field
\begin{equation}\label{3}
\ddot{\phi}+3H\dot{\phi}+\Gamma \dot{\phi} + m^2\phi=0,
\end{equation}
where $\Gamma $ is the rate of particle decay of the scalar field
into other particles. For example, if this scalar field decays in
two scalar fields $\phi \rightarrow \chi\chi$, the rate is
\cite{reheat}
\begin{equation}\label{4}
\Gamma \simeq \frac{g^4 \sigma^2}{8\pi m},
\end{equation}
where I am assuming a coupling $g^2\sigma\chi^2\phi$. The decay
products of the inflaton field are ultrarelativistic ($m\gg
m_\chi$), and their energy density decreases due to the expansion
of the universe much faster than the energy of the oscillating
field $\phi$. In this case reheating may {\it ends} when the Hubble
parameter $H\simeq 2/3t$ becomes smaller than $\Gamma$. However,
for $\sigma \ll \Phi$, where $\Phi$ is the amplitude of the oscillations,
we can write \cite{KLS97}: $\Gamma \simeq g^4\Phi^2/8\pi m$, then
because $\Phi^2$ decreases as $t^{-2}$ (see Eq.(\ref{2})) in the
expanding universe, whereas the Hubble parameter decreases only
as $t^{-1}$, the decay rate never catches up with the expansion of the
universe, and reheating never completes.

So even in the perturbative regime of reheating, we have a mechanism
to stop the process leaving part of the oscillating field $\phi$ decoupled
and behaving as dark matter. However, as we can see in the next sections,
this model by itself can not work, because it gives a result still in
contradiction with the observational constraints.

\section{Observational constraints}

In this section we derive a constraint for the initial amplitude
of the scalar field $\phi$ oscillations, which can be identified
as the dark matter component of the model. The idea is to compare
the theoretical results with an observational constraint.

As we saw in the previous section, during inflation the scalar
mass satisfy $m \ll H$, meanwhile for the quadratic potential the
condition $m \simeq H$ marks the end of inflation. To recover the
standard dark matter scenario, the scalar mass should satisfy $m
\gg H_{eq}$, where $H_{eq}$ is the value of the Hubble parameter
at the time of radiation and matter equality. If we denote by
$t^*$ the time at which the scalar mass equals the Hubble
parameter $m=H^{*}$, then
\begin{equation}\label{fried}
m^2=\frac{8\pi }{3M_p^2} \rho_{R}^*.
\end{equation}
We are assuming here that a large part of the scalar energy
density $\rho_{\phi}$ was transformed into radiation, dominating
the matter content of the universe, and the rest is still in the
form of an oscillating field. For $t>t^*$ the scalar field $\phi$
oscillate around the minimum of the potential. The energy density
average behaves as dust
\begin{equation}\label{rofi}
\rho_{\phi} = \frac{1}{2}m^2\phi{^2}_*\left(\frac{a_*}{a} \right)^3,
\end{equation}
and also the radiation component evolves
\begin{equation}\label{ror}
\rho_R=\rho_R^*\left(\frac{a_*}{a} \right)^4.
\end{equation}
What we want to compare with observations is the dark matter mass
per photon ratio $ \xi _{dm}=\rho_\phi/n_{\gamma}$. This quantity
is constant for $t>t^*$ apart from changes in the number of degrees
of freedom of the species. We assume here an adiabatic expansion
where $S=g_{S} T^3a^3$ is constant during the evolution, where
$g_{S}$ is the entropic degrees of freedom, usually very similar
to the total degrees of freedom $g_{*}$. Because
$n_\gamma=2\zeta(3)T^{3}/\pi^2$, then
\begin{equation}\label{numf}
n_{\gamma, 0}=n_{\gamma}^*
\frac{g_{*S}}{g_{S,0}}\left(\frac{a_{*}}{a_{0}}\right)^{3},
\end{equation}
where the zero subscript indicates current values. Notice that we
are assuming a change in the number of entropic degrees of
freedom. Then the current dark matter per photon ratio is
\begin{equation}\label{main}
\xi_{dm,0}=\frac{\rho_{\phi,0}}{n_{\gamma,0}} \simeq 4%
\frac{g_{S,0}}{g_{*}^{1/4}}\frac{m^{1/2}}{M_p^{1/2}}\frac{\phi^2_*}{M^2_p}M_p.
\end{equation}
The observational measure of this ratio is
$\xi_{dm,0}=2.2\times10^{-28}M_p$ using values from WMAP3 \cite{wmap}, which
for typical values $g_*\sim 100$, $g_{S,0}=3.9$ gives the following
constraint
\begin{equation}\label{vin}
\frac{m^{1/2}}{M_p^{1/2}}\frac{\phi^2_*}{M^2_p}\simeq 4\times
10^{-29}.
\end{equation}
Using considerations from structure formation, we get an upper
bound for the scalar field mass $m/M_p = 10^{-52}$ or
$m>10^{-23}$eV. Obtaining the correct amplitude for scalar
perturbations requires $m/M_p \simeq 10^{-6}$, then the condition
(\ref{vin}) imposes that $\phi_* \simeq  10^{-13}M_p$ which is in
contradiction with the initial statement for a value of $\phi \sim
M_p$ at the end of inflation. In this case to get an observable
viable model we need an incomplete reduction of the scalar field
amplitude during reheating \cite{LUL}, reducing the energy density
a factor of $10^{26}$.

As I stressed at the end of section II, even in the standard
picture of reheating we can have a partial decay of the inflaton
field. In this case, we find that at the beginning of the
oscillations the field amplitude is already $\sim 10^{-2}M_p$. So,
the constraint mentioned in the last paragraph implies that during
reheating the energy density must decays in $22$ orders of
magnitude.

\section{The problem}

So, what we need, is to have an incomplete reheating phase after
inflation, in agreement with the observational constraint: that of
having at the beginning of the oscillations (which is identified
with dark matter) an scalar field  amplitude $\phi_* \simeq 10^{-13}M_p$.
Clearly for a model where the amplitude of oscillations is very
large, we can not neglect the parametric resonant effects during
reheating \cite{preheat}.

Using {\it preheating}, the authors of \cite{LUL} found that, for
an interaction term $g^{2}\chi^{2}\phi^{2}$ where $g$ is the
coupling constant, the decay is incomplete, stoping until the
amplitude of the scalar field falls below $m/g$. Using the
required amplitude of the scalar field at the end of inflation
(see eq.(\ref{vin})), $\phi_{*}\simeq 10^{-7}m$ implies a coupling
constant value of $g \simeq 10^{7}$, which is clearly incompatible
with the model.

However the restriction can not be applied at the end of preheating,
because after the broad and narrow parametric resonance phases
\cite{preheat}, the reheating process follows through the standard
mechanism. So, these conditions have to be applied after the entire
reheating phase ends. In particular, the fact that the amplitude falls 
below $m/g$ only means that resonant production stops.

For a range $ g \sim 10^{-1} - 10^{-3}$ of the coupling constant, the
amplitude of the field after preheating is $\Phi \sim 10^{-5}M_p - 10^{-3}M_p$
respectively. These are the initial values for the standard reheating
phase, so the observational constraint implies that during particle decay
the energy density has to decrease 16 orders of magnitude and not 26 as
was settled in \cite{LUL}.

So, to get a viable observational model we have to consider the process
of reheating once the resonant phase has finished. In this new context, the
study of reheating is clearly different from the standard scenario; now the
universe not only has the contribution of the coherent oscillations of the inflaton
field but also the particles created during preheating. Therefore, our problem
is to find a mechanism that not only allows to avoid that the field inflaton decays
completely, but also take into account the presence of particles created in the
previous phase.

\section{Reheating and thermal masses}

As we can see below, both features are connected; the presence of
radiation during the perturbative reheating phase enable us to
stop the process. Let us assume that during the process of
preheating the inflaton decay products scatter and thermalize to
form a thermal background \cite{KNR}. This thermalized particle
species acquires a plasma mass $m_p (T)$ of the order of $\nu T$
where $\nu $ is the typical coupling governing the particle
interaction. The presence of thermal masses imply that the
inflaton zero mode cannot decay into light states if its mass $m$
is smaller than about $\nu T$. If we expressed the inflaton
zero-temperature decay width as $\Gamma_{\phi}=\alpha_{\phi} m$,
at finite temperature it becomes
\begin{equation}\label{gamat}
\Gamma_{\phi}(T)= \alpha_{\phi}m \sqrt{1-4\frac{\nu^2 T^2}{m^2}}.
\end{equation}
The system of equations to be solved is then
\begin{equation}\label{syseq}
\begin{array}{c}
          \dot{\rho_{\phi}}+ (3H + \Gamma_{\phi} )\rho_{\phi}=0, \\
          \dot{\rho_{R}}+ (4H\rho_{R} - \Gamma_{\phi} )\rho_{\phi}=0.
\end{array}
\end{equation}
When the plasma mass $m_p (T)=\nu T$ becomes comparable to the
inflaton mass $m$, the temperature reaches the value $T \simeq
m/2\nu$, remaining constant for a while, indicating that particle
creation stopped; $\Gamma_{\phi}=0$. At this time $\rho_R$ stays
constant and $\rho_{\phi}$ decays as $a^{-3}$. In the absence of
plasma masses, right after inflation the temperature increases
until $T_{max}$, and then decreases as $a^{3/8}$ until it reaches
the reheating temperature $T_{rh}$ \cite{ChKR}
\begin{equation}\label{trh}
T_{rh} = 0.2 \left(\frac{100}{g_{*}} \right)^{1/4} \alpha^{1/2} \sqrt{m M_p}.
\end{equation}
In the case $m < T_{rh}$, and using the values $m \simeq 10^{8}$GeV
and $g_{*}\sim 100$ (the effective number of relativistic degrees of
freedom) we obtain
\begin{equation}\label{alfa1}
\alpha_{\phi} \geq 3\times 10^{-10}.
\end{equation}
To estimate the duration of this $T$ constant phase, I use the fact that 
decays are not possible for $ T \geq m/2g$. The scalar field energy
density $\rho_{\phi} \simeq a^{-3}$ until it becomes smaller than $\rho_R$.
So the condition is simply
\begin{equation}\label{cond10}
V\left(\frac{a_i}{a_f} \right)^{3} \leq \frac{\pi^2}{30}g\left(\frac{m}{2g} \right)^4 
\end{equation}
where $a_i$ indicates the end of inflation and $a_f$ the time when the $\phi$ energy
density equals $\rho_R$. So, the universe evolves
\begin{equation}\label{afai}
\frac{a_f}{a_i}\simeq \left(\frac{V}{m^4} \right)^{1/3},
\end{equation}
which is equivalent to a number of e-foldings $N_e = 14 $ after
the standard particle decay process started, using the values
$V^{1/4} \sim 10^{13}$GeV and $m \sim 10^{8}$MeV \cite{KNR}. After
$\rho_R > \rho_{\phi}$, the universe becomes radiation dominated
and the rest of the inflaton energy density $\rho_{\phi}$ continue
its evolution as a coherently oscillating field, behaving as dark
matter. We can estimate the order of magnitude  of the decrease in
energy density using (\ref{afai})
\begin{equation}\label{rofroi}
\rho_{\phi}^f \simeq \rho_{\phi}^i \left(\frac{a_i}{a_f} \right)^{3} \sim \rho_{\phi}^i 10^{-18},
\end{equation}
which is in perfect agreement with the calculations of section IV.

\section{Summary}

In this paper, I have investigated the possibility to build a theoretical
model which describe both, the inflationary phase and also the dark
matter, using the inflaton field. The crucial ingredient is the period of
reheating after inflation, where almost all the energy stored in the inflaton
field is converted into relativistic particles. In fact, as the authors of
\cite{LUL} have stressed, the main condition that must be satisfied is
that not all the energy density of the inflaton decays into radiation. I
found that, even in the standard (perturbative) reheating scenario, the
process may end before all the inflaton energy density transformed into
radiation. However, this observation does not help to build the model by
itself, because for a large initial amplitude of the inflaton field (with a
typical value $\phi_i \sim M_p$ at the end of inflation), we can not neglect
the parametric resonance effects \cite{preheat}. Taking into account the
entire reheating phase, starting with the broad and narrow resonance
phases, and after that, the standard perturbative one, I found that this
can be accomplished. The key element in this description is that inflaton 
decay products acquire plasma masses \cite{KNR}, which may
cause inflaton decay to be kinematically forbidden, stopping the process.

\section*{Acknowledgments}

VHC was supported by DI-UNAB Grant 14-06/R.


\end{document}